\documentclass[a4paper,12pt]{article}
\pdfoutput=1
\usepackage{jcappub}             
\usepackage{amsmath,amssymb}
\usepackage{graphicx}
\usepackage[utf8]{inputenc}
\usepackage{cancel}
\usepackage[colorlinks=true]{hyperref}
\usepackage{color}
\usepackage{array,multirow}
\usepackage{subcaption}
\usepackage{float}
\usepackage{comment}
\usepackage{xcolor}

\usepackage{tikz} 
\usepackage{tikz-feynman}

\tikzfeynmanset{compat=1.1.0}
\usetikzlibrary{arrows,shapes}
\usetikzlibrary{trees}
\usetikzlibrary{matrix,arrows} 				% For commutative diagram
\usetikzlibrary{positioning}				% For "above of=" commands
\usetikzlibrary{calc,through}				% For coordinates
\usetikzlibrary{decorations.pathreplacing}  % For curly braces
\usepackage{pgffor}							% For repeating patterns

\usetikzlibrary{decorations.pathmorphing}	% For Feynman Diagrams
\usetikzlibrary{decorations.markings}
\tikzset{
    vector/.style={decorate, decoration={snake}, draw},
	provector/.style={decorate, decoration={snake,amplitude=2.5pt}, draw},
	antivector/.style={decorate, decoration={snake,amplitude=-2.5pt}, draw},
    fermion/.style={draw=black, postaction={decorate},
        decoration={markings,mark=at position .55 with {\arrow[draw=black]{>}}}},
    fermionbar/.style={draw=black, postaction={decorate},
        decoration={markings,mark=at position .55 with {\arrow[draw=black]{<}}}},
    fermionnoarrow/.style={draw=black},
    gluon/.style={decorate, draw=black,
        decoration={coil,amplitude=4pt, segment length=5pt}},
    scalar/.style={dashed,draw=black, postaction={decorate},
        decoration={markings,mark=at position .55 with {\arrow[draw=black]{>}}}},
    scalarbar/.style={dashed,draw=black, postaction={decorate},
        decoration={markings,mark=at position .55 with {\arrow[draw=black]{<}}}},
    scalarnoarrow/.style={dashed,draw=black},
    electron/.style={draw=black, postaction={decorate},
        decoration={markings,mark=at position .55 with {\arrow[draw=black]{>}}}},
	bigvector/.style={decorate, decoration={snake,amplitude=4pt}, draw},
    line/.style={draw=black},
}\usetikzlibrary{decorations.markings}

\usepackage[normalem]{ulem}

\usetikzlibrary{positioning}
\usetikzlibrary{decorations.pathmorphing}

\title{Two-Component Dark Matter in the Type-I 2HDM}
\author[a]{Patricio Escalona}
\author[b, c, d]{, Jacinto P. Neto}
\author[e]{, M. J. Neves}
\author[f, g]{, Camila Ramos}
\author[c]{, and David Suarez}

\affiliation[a]{Departamento de Física, Universidade Federal da Paraíba, 58051-970, João Pessoa, PB, Brazil}
\affiliation[b]{Departamento de F\'isica, Universidade Federal do Rio Grande do Norte, 59078-970, Natal, RN, Brazil}
\affiliation[c]{International Institute of Physics, Universidade Federal do Rio Grande do Norte, Campus Universit\'ario, Lagoa Nova, Natal-RN 59078-970, Brazil}
\affiliation[d]{University of Vienna, Faculty of Physics, Boltzmanngasse 5, A-1090 Vienna, Austria}
\affiliation[e]{Departamento de F\'isica, Universidade Federal Rural do Rio de Janeiro, BR 465-07, 23890-971, Serop\'edica, Rio de Janeiro, Brazil}
\affiliation[f]{Centro de Ci\^encias Naturais e Humanas, Universidade Federal do ABC, Santo Andr\'e, 09210-580
SP, Brazil}
\affiliation[g]{Instituto de F\'isica Te\'orica IFT-UAM/CSIC, Cantoblanco, E-28049, Madrid, Spain}

\emailAdd{jacinto.neto.100@ufrn.edu.br}
\emailAdd{ramos.camila@ufabc.edu.br}
\emailAdd{mariojr@ufrrj.br}
\emailAdd{patricioescalona96@gmail.com}
\emailAdd{david.suarezr@udea.edu.co}

\abstract{We investigate a two-component dark matter scenario in the type-I two-Higgs-doublet model. The dark sector contains a real scalar $s$ and a Dirac fermion $\chi$, whose stability is ensured by a $Z_4$ symmetry together with kinematic conditions. The scalar interacts with the visible sector through Higgs-portal couplings, while the fermion interacts with the scalar via Yukawa interactions. In this framework, we analyze the thermal freeze-out production of both candidates, accounting for annihilation, conversion, and semi-annihilation processes. A comprehensive scan over the multidimensional parameter space is performed in terms of physical masses, mixing angles, and portal couplings, imposing theoretical requirements such as perturbativity and vacuum stability. We confront the model with current experimental constraints, including the observed relic abundance, invisible Higgs decays, direct detection limits on spin-independent scattering cross sections, and electroweak precision observables. We find that viable regions of parameter space can satisfy all dark matter constraints, but collider bounds strongly constrain the scalar sector, narrowing the allowed regions and creating tension with those favored by dark matter phenomenology, particularly in the sub-TeV mass regime.
}

\begin{document}
\maketitle
\section{Introduction}

The Standard Model (SM) is remarkably successful but remains an incomplete theory because it fails to account for key phenomena such as neutrino oscillations~\cite{deSalas:2020pgw,Capozzi:2025wyn,Esteban:2024eli,ParticleDataGroup:2024cfk}, the baryon asymmetry of the Universe~\cite{Sakharov:1967dj,Bodeker:2020ghk}, and the nature of dark matter (DM)~\cite{Cirelli:2024ssz}. This cold, non-baryonic component accounts for about $25\%$ of the Universe's energy density. The DM problem is particularly compelling, given the robust astrophysical (e.g. galaxy rotation curves~\cite{1980ApJ...238..471R}), and cosmological (e.g. CMB fluctuations~\cite{Planck:2018vyg}) evidence for its existence and its central role in cosmic structure formation within the $\Lambda$CDM framework. Despite extensive experimental searches over recent decades, there is no conclusive non-gravitational detection of DM, and its microscopic nature remains unknown.

Thermal production in the early Universe made Weakly Interacting Massive Particles (WIMPs) leading candidates for DM. Although initial studies focused on ultraviolet complete models, giving rise to very strong candidates, such as nearly degenerate supersymmetric higgsinos~\cite{Chun:2016cnm}, the absence of weak-scale new physics shifted attention toward more phenomenological frameworks, including effective field theories~\cite{Criado:2021trs,Aebischer:2022wnl}, minimal SU$(2)_L$ multiplets \cite{Cirelli:2005uq,Cirelli:2009uv,Cirelli:2015bda,Saez:2018off,Belyaev:2018xpf,Cai:2017fmr,Acevedo:2024ava,Escalona:2024zkb}, and simplified models \cite{Arcadi:2024ukq,Berlin:2014tja,Baek:2015lna,Bell:2016uhg,ElHedri:2017nny,Jacques:2015zha,Xiang:2015lfa,Backovic:2015soa,Bell:2015rdw,Brennan:2016xjh}.
Simplified extensions of the SM with gauge singlets are particularly attractive for their simplicity and testable predictions, even though direct-detection limits strongly constrain them. To avoid these constraints, various mechanisms, including non-thermal DM production~\cite{Hall:2009bx}, and multi-component DM~\cite{PhysRevD.79.115002,Profumo:2009tb,Restrepo:2021kpq,Agudelo:2024luc,Qi:2026dwd,Qi:2025znq,Frank:2025jjt,Qi:2025jpm,Borah:2024emz,Ghorbani:2024jaa,YaserAyazi:2024dxk,Hosseini:2023qwu,Belanger:2022esk,Costa:2022oaa,Das:2022oyx,Ho:2022erb,DiazSaez:2021pfw} can be consistent with current observations, and they can generate distinctive experimental signatures compared to single-component scenarios.

Most studies of multi-component DM focus on scalar candidates, as the fermionic realizations typically require additional dark states or non-renormalizable couplings. In this work, we consider a framework containing a scalar and a fermionic singlet~\cite{DiazSaez:2021pmg, DiazSaez:2023wli}, extending a two-Higgs doublet model (2HDM), where a $Z_4$ symmetry stabilizes the DM~\cite{Cai:2015zza,Yaguna:2021rds}. Depending on the mass hierarchy and coupling strengths, the model features different production mechanisms, in which one or two singlets act as DM candidates.

The 2HDM is a well-motivated extension of the SM~\cite{Branco:2011iw}, and most SM ultraviolet extensions, such as supersymmetry, grand unified theories, or gauge extensions, include a 2HDM-like spectrum. This framework constitutes a diverse and active area of research~\cite{Tooby-Smith:2026kzj,Akamatsu:2026sjg,Meng:2026oil,Sah:2026aqr}. In particular, we consider a CP-conserving~\cite{Gunion:2002zf} type-I 2HDM~\cite{PhysRevD.41.3421} with two main dark sector interactions: Higgs-portal couplings linking the scalar to the two CP-even Higgs states, and Yukawa couplings connecting the scalar to the Dirac fermion. Our analysis incorporates the additional Higgs portal mediated by the CP-even partner, applies the latest direct detection constraints, and ensures consistency with electroweak precision tests. We also confront the model with scalar sector searches at colliders. These constraints allow us to understand the viable parameter space for two-component DM.

The paper is organized as follows: In \autoref{sec:model}, we present the essential features of the model, describe our treatment of the parameter space, and state the theoretical requirements of perturbativity and vacuum stability.
In \autoref{sec:genesis}, we discuss some mechanisms for DM generation and present the Boltzmann equations in the case where both particle densities are settled by thermal decoupling.
In \autoref{sec:pheno}, we consider observational and experimental constraints on the DM relic density, invisible decays of the Higgs boson, direct detection, and electroweak precision tests to perform a numerical analysis of the model's parameter space. We then discuss the tension of our scenario with collider results for the 2HDM.
Finally, in \autoref{sec:conclusions} we discuss the implications of the analysis for the model's viability and present our conclusions.
\section{The model}\label{sec:model}

The model investigated in this paper is an extension of the type-I 2HDM. This simple and well-motivated SM extension introduces a second $SU(2)_L$ scalar doublet with the same hypercharge as the SM Higgs doublet. We propose the addition of an inert real scalar $s$ and a vector-like Dirac fermion $\chi$. The Lagrangian reads 
\begin{eqnarray}\label{LDM}
{\cal L} \supset |D_{\mu}\Phi_1|^2+|D_{\mu}\Phi_{2}|^2+\frac{1}{2}(\partial_{\mu}s)^2+\overline{\chi}\left(i\gamma^{\mu}\partial_{\mu}-m_{\chi}\right)\chi
\nonumber \\
+\frac{1}{2}\left(  y_{s}\overline{\chi^{c}}\chi+y_{p}\overline{\chi^{c}}  \gamma_{5}  \chi+\mbox{H. c.}  \right)s-V(\Phi_{1},\Phi_{2},s) \; ,
\end{eqnarray}
where $\Phi_{i}$ are two scalar doublets of $SU(2)_L$ whose components are parametrized as
\begin{eqnarray}
\Phi_i=
\left(
\begin{array}{c}
\phi_{i}^{+} \\
\phi_i^0  \\
\end{array}
\right), \quad i=1,2 \; ,
\end{eqnarray}
the scalar ($y_{s}$) and pseudoscalar ($y_{p}$) Yukawa couplings mediate dark sector interactions, which can be made real by field rotations\footnote{If $\chi$ is a Majorana fermion, the relative phase of $y_s$ and $y_p$ is physical and leads to CP violation.}, and the scalar potential is given by
\begin{eqnarray}\label{Vsc}
    V(\Phi_1,\Phi_{2},s) &=& \mu_1^{ 2} (\Phi_{1}^{\dagger} \Phi_1) +\mu_2^{ 2} (\Phi_{2}^{\dagger} \Phi_2) -(\mu_{12}^{2}\Phi_{1}^{\dagger}\Phi_2+\mbox{H. c.}) \nonumber \\
    &&
    +\frac{\lambda_1}{2} (\Phi_1^{\dagger} \Phi_1)^{2} +\frac{\lambda_2}{2} (\Phi_2^{\dagger} \Phi_2)^{2} +\lambda_{3}(\Phi_1^{\dagger} \Phi_1)(\Phi_2^{\dagger} \Phi_2) \nonumber \\
    && +\lambda_{4}(\Phi_1^{\dagger} \Phi_2)(\Phi_2^{\dagger} \Phi_1) +\frac{\lambda_5}{2} \left[(\Phi_1^{\dagger} \Phi_2)^2+(\Phi_2^{\dagger} \Phi_1)^2  \right] \nonumber \\
    && +\frac{\mu_{s}^{ 2}}{2}  s^2 +\frac{\lambda_{s}}{4}  s^{4} + \frac{\lambda_{1s}}{2}  |\Phi_{1}|^{2}  s^{2}+\frac{\lambda_{2s}}{2}  |\Phi_{2}|^{2}  s^{2} \; ,
\end{eqnarray}
which respects the SM gauge symmetry group, $SU(3)_C \times SU(2)_L \times U(1)_Y$. In addition, we also impose that the fields transform under two discrete symmetries $Z_2^{\rm 2HDM-I}$ and $Z_4^\text{DM}$. 

The role of the $Z_2^{\rm 2HDM-I}$ symmetry is to avoid flavor-changing neutral currents (FCNCs) in the Yukawa interactions of the Higgs doublets with SM fermions, following the Paschos-Weinberg-Salam theorem\footnote{The $Z_2$ charge assignment to avoid FCNC is not unique and leads to the different types of 2HDM.}~\cite{Paschos:1976ay,Glashow:1976nt}.
The $Z_4^\text{DM}$ symmetry ensures the stability of the fermion $\chi$ and the scalar $s$, provided that the decay channel $s\to\bar{\chi}\chi$ is kinematically forbidden, i.e. requiring $m_{s} < 2 \, m_{\chi}$. Under these assumptions, both $s$ and $\chi$ can be simultaneously DM candidates.
Explicitly, the non-trivial transformation properties of the fields under these discrete symmetries are
\begin{subequations}
\begin{align}
    Z_{2}^{\text{2HDM-I}} &: \quad  \Phi_{1} \;\longrightarrow\; -\,\Phi_{1} \,, \\[6pt]
    Z_{4}^{\text{DM}}     &: \quad  
        s \;\longrightarrow\; -\,s \,, 
        \qquad 
        \chi \;\longrightarrow\; i\,\chi \,.
\end{align}
\end{subequations}
Notably, in \autoref{Vsc}, the $Z_{2}^{\text{2HDM-I}}$ is softly broken by the $\mu_{12}$ term, avoiding the domain wall problem~\cite{Zeldovich:1974uw}. 
Moreover, the $Z_{2}^{\text{2HDM-I}}$ forces the Yukawa interactions to be of the form
\begin{eqnarray}
- \, \mathcal{L} _{Y _{\text{2HDM}}} = y^{d}_{ij} \, \overline{Q} _{iL} \, \Phi _2 \, d_{jR} 
+ y^{u}_{ij} \, \overline{Q}_{iL} \, \widetilde{\Phi}_2 \, u_{jR} + y^{e}_{ij} \, \overline{L} _{iL} \, \Phi_2 \, e_{jR} + {\rm H. c.} \; ,  
\end{eqnarray}
where $i,j=1,2,3$ are family indices, $y^{d}_{ij}$, $y^{u}_{ij}$, $y^{e}_{ij}$ are Yukawa couplings and  $\widetilde{\Phi}_2=i\sigma_{2}\,\Phi_{2}^{\ast}$ is the isospin conjugate of $\Phi_2$.
The fermion content is the same as in the SM:
$Q_{iL}=\left(\, u_{iL} \;\; d_{iL} \, \right)^{T}$ is the left-handed quark doublet, $L_{iL}=\left(\, \nu_{iL} \;\; e_{iL} \, \right)^{T}$ is the left-handed lepton doublet, $u_{iR}$ and $d_{iR}$ are the right-handed quark singlets, and $e_{iR}$ are the right-handed lepton singlets. In this scenario, SM fermions interact only with $\Phi_{2}$, which provides the mass term for such fermions after electroweak symmetry breaking (EWSB). 
%These terms lead to fermion masses after spontaneous symmetry breaking.
%The electric charges of the particle content satisfy the usual relation $Q=I_{3L}+Y$, where $I_{3L}=\pm 1/2$ is the isospin quantum number for doublets, and $Y$ is the hypercharge.

The scalar spectrum after EWSB is obtained from the potential in \autoref{Vsc}, by shifting the neutral fields as
\begin{equation}\label{Phi1Phi2}
    \phi_i^{0}=\frac{1}{\sqrt{2}}\left(v_i+\rho_i+i\eta_i\right) \; .
\end{equation}
Since $s$ does not develop a vacuum expectation value (VEV), the minimization of the potential using the parametrization in \autoref{Phi1Phi2} leads to the following minimization conditions
%We consider $s$ an inert field which does not develop a VEV. When we minimize the potential and use the parameterization in \autoref{Phi1Phi2}, we obtain the minimum conditions
%
\begin{subequations}\label{eq:minimum}
\begin{eqnarray}
\mu_1^2 \, v_1-\mu_{12}^2 \, v_2+\frac{\lambda_1}{2} \, v_1^3+\frac{1}{2}\,\lambda\,v_1\,v_2^2 =0 \; ,\\ 
\mu_2^2 \, v_2-\mu_{12}^2 \, v_1+\frac{\lambda_2}{2} \, v_2^3+\frac{1}{2}\,\lambda\,v_1^2\,v_2
=0 \; ,
\end{eqnarray}
\end{subequations}
where we define $\lambda\equiv\lambda_{3}+\lambda_{4}+\lambda_5$.
In Appendix \ref{App:matrices}, we present the definitions of mass matrices, eigenstates, and mixing angles in more detail. Here, we show the spectrum of the theory after EWSB. First, the gauge bosons $Z$ and $W^\pm$ acquire masses
\begin{equation}
m_W=\frac{g}{2} \, \sqrt{v_1^2+v_2^2}
\hspace{0.5cm} \mbox{and} \hspace{0.5cm}
m_Z=\frac{m_W}{\cos \theta_W} \; ,
\end{equation}
where $g$ is the SU$(2)_L$ coupling, and $\theta_W$ is the Weinberg angle. To recover the measured values of these masses, the SM VEV is given by% we impose the relation
\begin{equation}\label{eq:sm_vev}
v \equiv \sqrt{v_1^2+v_2^2}=246\text{ GeV} \; .
\end{equation}
The scalar spectrum includes two CP-odd scalars. One is the Goldstone boson that becomes the longitudinal component of the neutral gauge boson $Z$. The other is a massive physical pseudoscalar $A$ with mass
\begin{equation}\label{eq:Amass}
m_A^2 = v^2\left(\frac{\mu^2_{12}}{v_1v_2}-\lambda_5\right) \; .
\end{equation}
There are also two pairs of charged scalars. One pair corresponds to the Goldstone bosons that become the longitudinal components of $W^{\pm}$. The other pair is a massive charged scalar $h^{\pm}$ with mass
\begin{equation}\label{eq:hpmass}
m_{h^\pm}^2 = v^2 \left[\frac{\mu^2_{12}}{v_1v_2}-\frac{1}{2}(\lambda_4-\lambda_5)\right] \; .
\end{equation}
Additionally, the scalar DM mass is 
\begin{eqnarray}
m_{s}^2=\mu_s^2+\frac{1}{2} \, \lambda_{1s}\,v_1^2+\frac{1}{2} \, \lambda_{2s}\,v_2^2 \; .
\end{eqnarray}
A single angle $\beta$ parametrizes the rotation matrix that diagonalizes both the CP-odd sector and the charged sector mass matrices. That angle $\beta$ satisfies
\begin{equation}\label{eq:betadef}
    \tan \beta = \frac{v_2}{v_1} \; .
\end{equation}
Finally, the model comprises two CP-even scalars with masses
\begin{equation}\label{h_and_H}
m_{h,H}^2 = \frac{1}{2}\left[\frac{\mu_{12}^2v^2}{v_1v_2}+\lambda_1v_1^2+\lambda_2v_2^2 \mp \sqrt{\left(\frac{\mu_{12}^2\left(v_1^2-v_2^2\right)}{v_1v_2}-\lambda_1v_1^2+\lambda_2v_2^2\right)^2+4\left(\mu_{12}^2-\lambda v_1 v_2\right)^2}\right] \; .
\end{equation}
On the other hand, a single angle $\alpha$ parametrizes the rotation matrix that diagonalizes the CP-even sector mass matrix, and this angle $\alpha$ satisfies
\begin{equation}\label{eq:alphadef}
\tan 2\alpha = \frac{2\lambda v_1^2v_2^2 -2\mu_{12}^2v_1v_2}{\mu_{12}^2 (v_2^2-v_1^2)+\lambda_1v_1^3v_2-\lambda_2v_1v_2^3} \; .
\end{equation}
Notice that, by construction, $h$ is lighter than $H$, and one of them must resemble the SM-like 125 GeV Higgs boson. Therefore, we distinguish two regimes: (i) when the mass of $h$ is $m_h=125$ GeV and $H$ is a heavy partner with mass $m_H>125$ GeV, and (ii) when the mass of $H$ is $m_H = 125$ GeV and $h$ is a lighter partner with mass $m_h < $ 125 GeV. 

It is important to note that for arbitrary mixing angles $\alpha$ and $\beta$, mass conditions alone are insufficient to guarantee that the $125$ GeV Higgs boson replicates the properties of the SM Higgs. The {\it alignment limit} denotes specific configurations of these mixing angles in which we recover the couplings and behavior of the SM-like Higgs boson. For the regime (i), this corresponds to $\cos{(\alpha-\beta)}=1$, while for regime (ii), the corresponding configuration is $\sin{(\alpha-\beta)}=1$. Depending on the regime, we determine the interactions of CP-even scalars with SM fermions and gauge bosons~\cite{Carena:2013ooa}. In this work, we conservatively consider the {\it exact} alignment limit. 

We show the interactions of $s$ with CP-even scalars in the last line of \autoref{Vsc}. By using \autoref{Phi1Phi2}, which correspond to
\begin{equation}
{\cal L}_{sc}^{int}=\frac{\lambda_{s}}{4} \, s^{4} +\frac{\lambda_{1s}}{2} \, s^{2} \left( v_1\,\rho_1+\frac{\rho_1^2}{2} \right) +\frac{\lambda_{2s}}{2} \, s^{2} \left( v_2\,\rho_2+\frac{\rho_2^2}{2}\right) \; .
\end{equation}
We rotate the fields into physical ones, via \autoref{transfh}. Consequently, the interaction terms of the scalar DM $s$ with $h$ and $H$ are 
\begin{eqnarray}
{\cal L}_{sc}^{int}&=&\frac{\lambda_{s}}{4} \, s^{4}
+\frac{1}{2} \, \lambda_{sh} \, v \, s^2 \, h
+\frac{1}{2} \, \lambda_{sH} \, v \, s^2 \, H
\nonumber \\
&&
+\frac{1}{4} \left(\lambda_{1s}\cos^2\alpha+\lambda_{2s}\sin^2\alpha \right) s^2 \, h^2
\nonumber \\
&&
+\frac{1}{4} \, \sin(2\alpha) \left(\lambda_{2s}-\lambda_{1s}\right) s^2 \, H \, h
+\frac{1}{4} \left(\lambda_{1s}\sin^2\alpha+\lambda_{2s}\cos^2\alpha \right) s^2 \, H^2 \; ,
\end{eqnarray}
where we define the couplings
\begin{subequations}
\begin{eqnarray}
\lambda_{sh}=\lambda_{1s} \, \cos\beta \, \cos\alpha+\lambda_{2s} \, \sin\beta \, \sin\alpha\,, \\
\lambda_{sH}=\lambda_{2s} \, \cos\beta \, \cos\alpha-\lambda_{1s} \, \sin\beta \, \sin\alpha \; ,
\end{eqnarray}
\end{subequations}
which are the effective portals between the scalar DM and the SM through the two Higgs-like states. The fermion DM lacks direct interactions with the SM fields.

We now describe how we handle the model's parameter space. The potential of the type-I 2HDM has 8 parameters, namely 
\begin{equation}
\{\lambda_1,\lambda_2,\lambda_3,\lambda_4,\lambda_5,\mu_1,\mu_2,\mu_{12}\} \; ,    
\end{equation}
while our extension introduces 7 more: 
\begin{equation}
\{\lambda_{s},\lambda_{1s}, \lambda_{2s},y_s,y_p,\mu_s,m_\chi\} \; .    
\end{equation}
To perform the numerical scan, we work with the relevant free parameters, namely the physical masses, mixing angles, and couplings~\cite{Gunion:2002zf}:
\begin{equation}\label{eq:inputpar}
\{m_h, m_H, m_A, m_{h^{+}}, \beta, \sin(\beta-\alpha), m_s,m_\chi, \lambda_{1s}, \lambda_{2s},y_L,y_R\} \; . 
\end{equation}
%we find it inconvenient to use the original 2HDM Lagrangian parameters as independent parameters, so we treat them as functions of the physical masses, mixing angles, and the SM VEV. Explicitly, we choose the following parameters as independent input:
%
%The Lagrangian parameters are related to these by: Equations D.(13-17) of Ref.~\cite{Gunion:2002zf}, the minimum conditions~\autoref{eq:minimum}, and the mixing angle definitions, \autoref{eq:betadef} and \autoref{eq:alphadef}. %\mu_{12} 
%We fix the mass $m_h$ ($m_H$) to be $125$ GeV for the regimes with heavy (light) Higgs partner, and following the alignment limit, we also fix the value of $\sin(\beta-\alpha)$ to be $0$ or $1$ accordingly. In the dark sector, we choose 
%
%\begin{equation}\label{eq:inputDM}
%\{m_s,m_\chi,\lambda_s,\lambda_{1s}, \lambda_{2s},y_L,y_R\} \; ,
%\end{equation}
%
We fix the mass $m_h$ ($m_H$) to be $125$ GeV for the regimes with heavy (light) Higgs partner, and following the alignment limit, we also fix the value of $\sin(\beta-\alpha)$ to be $0$ or $1$ accordingly. Furthermore, $y_L=y_s+y_p$ and $y_R=y_s-y_p$ as defined by the Weyl decomposition of the Dirac fermion $\chi$. Note that $\lambda_s$ modulates the scalar DM self-interactions and is irrelevant for DM phenomenology and the scalar spectrum\footnote{Although this parameter can have an impact on structure formation~\cite{Tulin:2017ara,Harvey:2015hha,Sagunski:2020spe,Bondarenko:2017rfu,DES:2023bzs,Adhikari:2022sbh} and receive Bullet cluster constraints~\cite{Clowe:2006eq}.}.
In \autoref{sec:pheno}, we perform a numerical scan over this 12-dimensional parameter space.

%%%%% theoretical constraints
\subsection{Theoretical constraints}
Here, we present two theoretical conditions that are critical to the consistency of our calculations. First, we require the quartic and Yukawa couplings to lie in the perturbative regime,
\begin{equation}
|\lambda_i| \; , \; |y_s| \; , \; |y_p| \leq 4\pi, \; \mbox{with} \; i=1,2,3,4,5 \; ,
\end{equation}
which keeps loop corrections subdominant compared with tree-level effects.Additionally, the vacuum stability conditions for the scalar potential read
\begin{subequations}
\begin{eqnarray}
\lambda_1\geq 0 \, ,\quad \lambda_2\geq 0 , , \quad  \lambda_s\geq 0\, , \\
\sqrt{2\lambda_1 \lambda_s}+2 \lambda_{1s}\geq 0, \quad \sqrt{2\lambda_2 \lambda_s}+2 \lambda_{2s}\geq 0 \; .
\end{eqnarray}
\end{subequations}
In addition to these inequalities, when  $\lambda_4 \geq \lambda_5$, the following is required:
\begin{equation}
\frac{\sqrt{\lambda_1 \lambda_2}}{2}+ \lambda_3\geq 0 \; ,
\end{equation}
and
\begin{align}
\frac{1}{2\sqrt{2}}\sqrt{\left(\sqrt{2} \sqrt{\lambda_1 \lambda_s}+2 \lambda_{1s}\right) \left(\frac{\sqrt{\lambda_1 \lambda_2}}{2}+\lambda_3\right) \left(\sqrt{2} \sqrt{\lambda_2\nonumber
   \lambda_s}+2 \lambda_{2s}\right)}\\+\frac{1}{4} \sqrt{\lambda_1 \lambda_2 \lambda_s}+\frac{\sqrt{\lambda_1} \lambda_{2s}}{2 \sqrt{2}}+\frac{\lambda_{1s}
   \sqrt{\lambda_2}}{2 \sqrt{2}}+\frac{\lambda_3 \sqrt{\lambda_s}}{2}\geq 0 \; .
\end{align}
while, if $\lambda_4 < \lambda_5$, the inequalities are as follows:
\begin{equation}
\frac{\sqrt{\lambda_1 \lambda_2}}{2}+ \lambda_3 + \lambda_4-\lambda_5\geq 0 \; ,
\end{equation}
and
\begin{align}
\sqrt{\left(\sqrt{2} \sqrt{\lambda_1 \lambda_s}+2
\lambda_{1s}\right) \left(\sqrt{2} \sqrt{\lambda_2 \lambda_s}+2
   \lambda_{2s}\right) \left(\sqrt{\lambda_1 \lambda_2}+2
   (\lambda_3+\lambda_4-\lambda_5)\right)} 
   \nonumber \\ 
   +\sqrt{\lambda_1 \lambda_2
   \lambda_s}+\sqrt{2} \sqrt{\lambda_1} \lambda_{2s}+\sqrt{2} \lambda_{1s}
   \sqrt{\lambda_2}+2 \sqrt{\lambda_s}
   (\lambda_3+\lambda_4-\lambda_5)\geq 0 \; .
\end{align}
\section{Dark matter production}
\label{sec:genesis}

The DM production mechanism strongly depends on the couplings between the 2HDM sector and the dark sector, $\lambda_{1s}$ and $\lambda_{2s}$, as well as the couplings within the dark sector, $y_s$ and $y_p$. Additionally, the mass hierarchy between $m_\chi$ and $m_s$ also strongly impacts the DM production. 
Higgs portal values in the range $\max{(\lambda_{1s},\lambda_{2s})\sim10^{-12}-10^{-6}}$ do not bring the dark sector into thermal equilibrium with the rest of the plasma, and DM production may proceed through freeze-in or dark freeze-out mechanisms~\cite{Hall:2009bx,Bernal:2017kxu}. Conversely, when at least one of these couplings is sufficiently large, the singlet scalar reaches thermal equilibrium with the 2HDM. In this regime, the dark sector coupling naturally fall into two distinct possibilities:
\begin{itemize}
    \item $\max{(|y_s|,|y_p|)} \lesssim 10^{-6}$: In this case, the two DM components feebly interact with each other. If kinematically allowed ($m_s > 2m_\chi$), the scalar $s$ decays into $\chi$ pairs and, together with $2 \to 2$ scattering processes, leads to freeze-in production of $\chi$. In this case, $s$ becomes unstable and does not contribute to the present dark matter abundance. On the other hand, for $m_s < 2m_\chi$, the phenomenology effectively reduces to that of a 2HDM extended by a scalar singlet~\cite{Boucenna:2011hy}, where $s$ is produced via freeze-in while the fermion $\chi$ neither thermalizes nor attains a significant relic density, remaining phenomenologically irrelevant.
    \item $\max{(|y_s|,|y_p|)} \gtrsim 10^{-6}$: In this case, both $s$ and $\chi$ reach thermal equilibrium, and the freeze-out mechanism determines their relic density. In what follows, we focus on this coupling regime with $m_s < 2m_\chi$.
\end{itemize}

\subsection{Freeze-out production}
\begin{figure}[t!]
\centering
\begin{tikzpicture}[line width=1.0 pt, scale=0.55]
\begin{scope}[shift={(0,5)}]
	\draw[fermion](-3,1) -- (-1,0);
	\draw[fermionbar](-3,-1) -- (-1,0);
	\draw[scalarnoarrow](-1,0) -- (1,0);
	\draw[scalarnoarrow](1,0) -- (3,1);
	\draw[scalarnoarrow](1,0) -- (3,-1);
    \node at (-3.5,1.0) {$\chi$};
	\node at (-3.5,-1.0) {$\bar{\chi}$};
    \node at (0,0.5) {$s$};
    \node at (1.7,1) {$s$};
    \node at (1.8,-1.0) {$\varphi$};
    \node at (0,-3) {$\textit{(a)}$};
\end{scope}
\begin{scope}[shift={(8,6)}]
	\draw[fermion](-3,1) -- (-1,0);
	\draw[fermion](-1,0) -- (1,1);
	\draw[scalarnoarrow](-1,0) -- (-1,-2);
	\draw[scalarnoarrow](-1,-2) -- (-3,-3);
	\draw[scalarnoarrow](-1,-2) -- (1,-3);
    \node at (-3.5,1.0) {$\chi$};
	\node at (1.5,1.0) {$\chi$};
    \node at (-1.5,-1) {$s$};
	\node at (-3.5,-2.8) {$s$};
    \node at (1.5,-2.8) {$\varphi$};
    \node at (-0.9,-4) {$\textit{(b)}$};
\end{scope}
\begin{scope}[shift={(15,5)}]
    \draw[fermion](-3,1) -- (-1,1);
	\draw[fermionbar](-3,-1) -- (-1,-1);
	\draw[fermion](-1,1) -- (-1,-1);
	\draw[scalarnoarrow](-1,1) -- (1,1);
	\draw[scalarnoarrow](-1,-1) -- (1,-1);
    \node at (-3.6,1.0) {$\chi$};
	\node at (-3.6,-1.0) {$\chi$};
    \node at (-1.6,0) {$\chi$};
	\node at (1.4,1) {$s$};
    \node at (1.4,-1) {$s$}; 
    \node at (-0.8,-3) {$\textit{(c)}$};
\end{scope}
\begin{scope}[shift={(22,5)}]
	\draw[fermion](-3,1) -- (-1,1);
	\draw[fermionbar](-3,-1) -- (-1,-1);
	\draw[fermion](-1,1) -- (-1,-1);
	\draw[scalarnoarrow](-1,1) -- (1,-1);
	\draw[scalarnoarrow](-1,-1) -- (1,1);
    \node at (-3.6,1.0) {$\chi$};
	\node at (-3.6,-1.0) {$\chi$};
    \node at (-1.6,0) {$\chi$};
	\node at (1.4,1) {$s$};
    \node at (1.4,-1) {$s$};
    \node at (-0.8,-3) {$\textit{(d)}$};
\end{scope}
\begin{scope}[shift={(0,-1)}]
	\draw[scalarnoarrow](-2.5,1) -- (-1,0);
	\draw[scalarnoarrow](-2.5,-1) -- (-1,0);
	\draw[scalarnoarrow](-1,0) -- (1,0);
	\draw[line](1,0) -- (2.5,1);
	\draw[line](1,0) -- (2.5,-1);
    \node at (-3,1.0) {$s$};
	\node at (-3,-1.0) {$s$};
    \node at (-0.1,0.46) {$\varphi$};
	\node at (1.4,1.3) {$X$};
    \node at (1.4,-1.3) {$X$};
    \node at (-0,-2.5) {$\textit{(e)}$};
\end{scope}
\begin{scope}[shift={(8,-1)}]
	\draw[scalarnoarrow](-3,1) -- (-1,1);
	\draw[scalarnoarrow](-3,-1) -- (-1,-1);
	\draw[scalarnoarrow](-1,1) -- (-1,-1);
	\draw[scalarnoarrow](-1,1) -- (1,1);
	\draw[scalarnoarrow](-1,-1) -- (1,-1);
    \node at (-3.5,1.0) {$s$};
	\node at (-3.5,-1.0) {$s$};
    \node at (-1.5,0) {$s$};
	\node at (1.5,1) {$\varphi$};
    \node at (1.5,-1) {$\varphi$};
    \node at (-0.8,-2.5) {$\textit{(f)}$};
\end{scope}
\begin{scope}[shift={(15,-1)}]
	\draw[scalarnoarrow](-3,1) -- (-1,1);
	\draw[scalarnoarrow](-3,-1) -- (-1,-1);
	\draw[scalarnoarrow](-1,1) -- (-1,-1);
	\draw[scalarnoarrow](-1,1) -- (1,-1);
	\draw[scalarnoarrow](-1,-1) -- (1,1);
    \node at (-3.4,1.0) {$s$};
	\node at (-3.4,-1.0) {$s$};
    \node at (-1.5,0) {$s$};
	\node at (1.5,1) {$\varphi$};
    \node at (1.5,-1) {$\varphi$};
    \node at (-0.8,-2.5) {$\textit{(g)}$};
\end{scope}
\begin{scope}[shift={(22,-1)}]
	\draw[scalarnoarrow](-2.5,1) -- (-1,0);
	\draw[scalarnoarrow](-2.5,-1) -- (-1,0);
	\draw[scalarnoarrow](-1,0) -- (0.5,1);
	\draw[scalarnoarrow](-1,0) -- (0.5,-1);
    \node at (-3,1.0) {$s$};
	\node at (-3,-1.0) {$s$};
	\node at (1,1) {$\varphi$};
    \node at (1,-1) {$\varphi$};    
    \node at (-0.8,-2.5) {$\textit{(h)}$};
\end{scope}
\end{tikzpicture}
\caption{Relevant diagrams for freeze-out. Here, $\varphi$ collectively denotes the CP-even scalars $h$ and $H$, while $X$ represents SM particles and additional 2HDM scalars ($h, H, h^\pm, A$). Diagrams (\textit{a}) and (\textit{b}) correspond to $s$- and $t$-channel semi-annihilation processes, respectively; (\textit{c}) and (\textit{d}) to DM conversion processes; and (\textit{e})--(\textit{h}) to $s$ pair annihilation into $X$ final states.}\label{diagrams}
\end{figure}
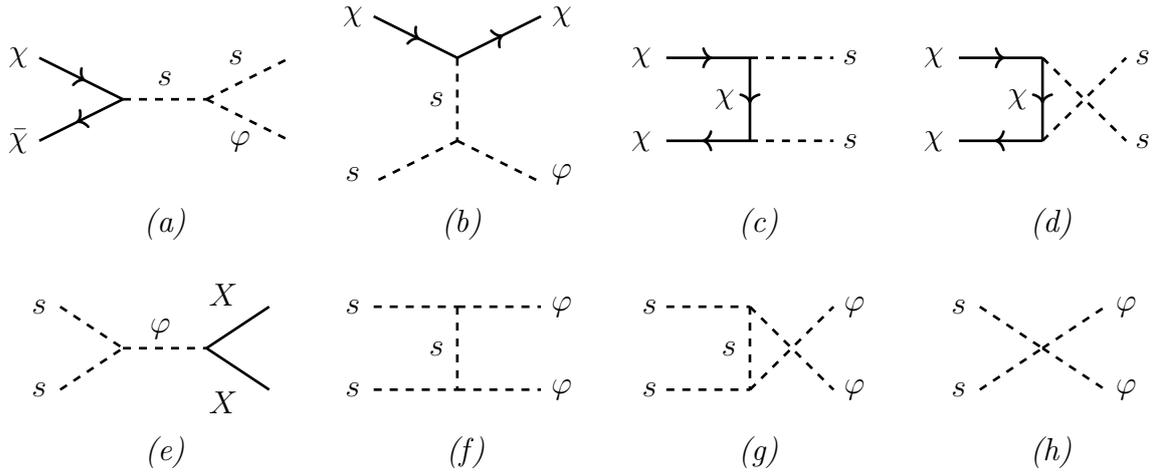
As the Universe expands and cools, both DM particles thermally decouple, and the rates of the processes induced by the diagrams shown in \autoref{diagrams} determine their relic abundance. Semiannihilation processes (diagrams (a)–(b))  naturally occur in DM models with stabilizing discrete symmetries larger than $Z_2$.  Renormalizable theories with fermionic semi-annihilating DM require bosonic degrees of freedom in the dark sector for such interactions. In our setup, this boson is a DM particle itself, which generate conversion processes (diagrams (c)–(d)) involving two DM particles in the final state. Finally, the standard annihilation channels (diagrams (e)-(h)), in which no DM particle appears in the final state, are also relevant for determining the relic abundance.
The evolution of the yield  of the $i$-particle species in the plasma, $Y_i\equiv n_i/s$, where $n_i$ is the number density of DM, and $s$ is the entropy density, as a function of the ratio $x\equiv\mu/T$, where $\mu$ is the reduced mass of the DM system, $\mu=m_s m_\chi/(m_s+m_\chi)$, and $T$ is the photon temperature, is governed by the  following set of coupled Boltzmann equations:
\begin{subequations}
\begin{eqnarray}\label{boltzc}
 \frac{dY_\chi}{dx} &=& - \lambda_{\chi\bar{\chi}ss}\left(Y_\chi^2 - Y_s^2\frac{Y_{\chi,e}^2}{Y_{s,e}^2}\right) - \lambda_{\chi\bar{\chi}s\varphi}\left(Y_\chi^2 - Y_s\frac{Y_{\chi ,e}^2}{Y_{s,e}}\right) \, , 
 \\
 \frac{dY_s}{dx} &=& - \lambda_{ssXX} \left(Y_s^2 - Y_{s,e}^2\right) + \lambda_{\chi\bar{\chi}ss} \left(Y_\chi^2 - Y_s^2\frac{Y_{\chi ,e}^2}{Y_{s,e}^2}\right) \nonumber\\ 
& + & \frac{1}{2}\lambda_{\chi\bar{\chi}s\varphi}\left(Y_\chi^2 - Y_s\frac{Y_{\chi ,e}^2}{Y_{s,e}}\right) -\frac{1}{2} \lambda_{s\chi\chi \varphi}\left(Y_sY_\chi - Y_\chi Y_{s,e}\right),
\end{eqnarray}
\end{subequations}
where we define
\begin{align}
	\lambda_{ijkl}(x)\equiv
	\dfrac{\langle\sigma_{ijkl}v\rangle(x)\cdot s(T)}
	{x\cdot H(T)} \; , \qquad \text{for }  i,j,k,l=\chi,s,\varphi,X.
	\label{def_lambda}
	\end{align}
where $\varphi$ is either $h$ or $H$, $X$ denotes a 2HDM particle (SM particles plus the additional scalars in the visible sector), and $\langle\sigma v\rangle$ is the thermally averaged cross-section times velocity. The entropy density and Hubble rate in a radiation-dominated Universe are functions of the temperature:
\begin{align}\label{Hubbleandentropy}
H(T)=\sqrt{\dfrac{4\pi^3G}{45}g_{*}(T)} \cdot T^2, \quad 
	s(T)=\dfrac{2\pi^2}{45} \, g_{*s}(T)\cdot T^3 \; ,
\end{align}
where $G$ is the Newton gravitational constant, and $g_*$ and $g_{s*}$ are the effective degrees of freedom that contribute, respectively, to the energy and the entropy density \cite{Husdal:2016haj}. The equilibrium densities, $Y_{i,e} \equiv n_{i,e}/s$, are calculated using the Maxwell-Boltzmann distribution
\begin{align}\label{eqden}
n_{i,e}(T) = g_i \, \dfrac{m_i^2}{2\pi^2} \, T \, K_2\left(\frac{m_i}{T}\right) \; ,
\end{align}
with $g_i$ being the internal degrees of freedom, and $K_2$ is the modified Bessel function of the second kind. We use the detailed balance principle
\begin{equation}
n_{a,e} \, n_{b,e}  \langle \sigma_{abcd} v\rangle 
=  n_{c,e} \, n_{d,e} \langle \sigma_{cdab} v\rangle \; , 
\end{equation}
to modify the above equations as the mass hierarchies change. 

\begin{figure}
    \centering
    \includegraphics[width=0.49\linewidth]{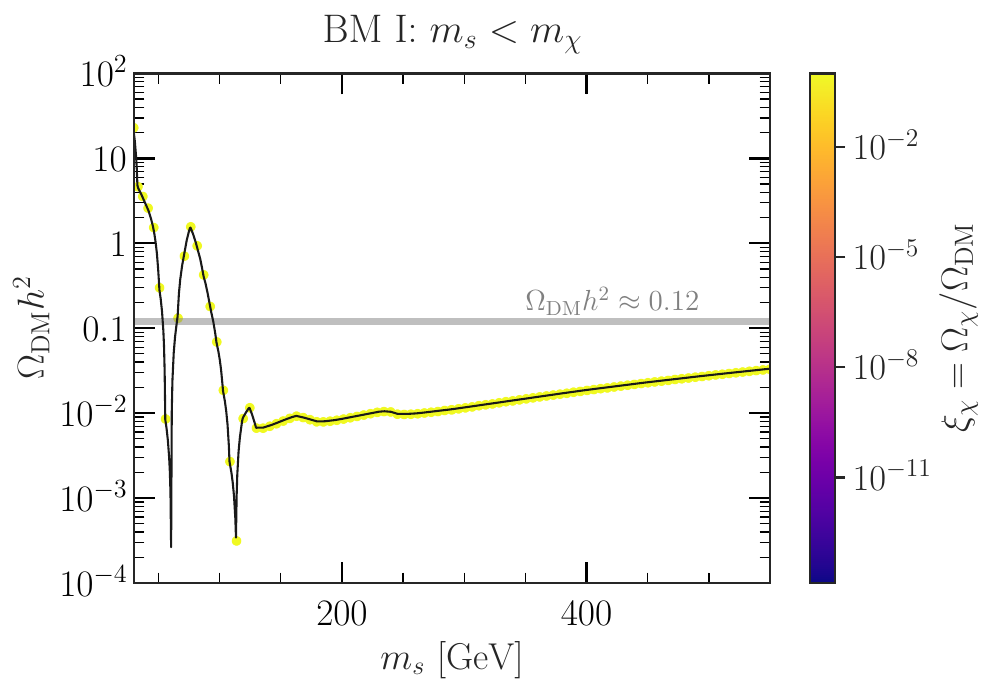}
    \includegraphics[width=0.49\linewidth]{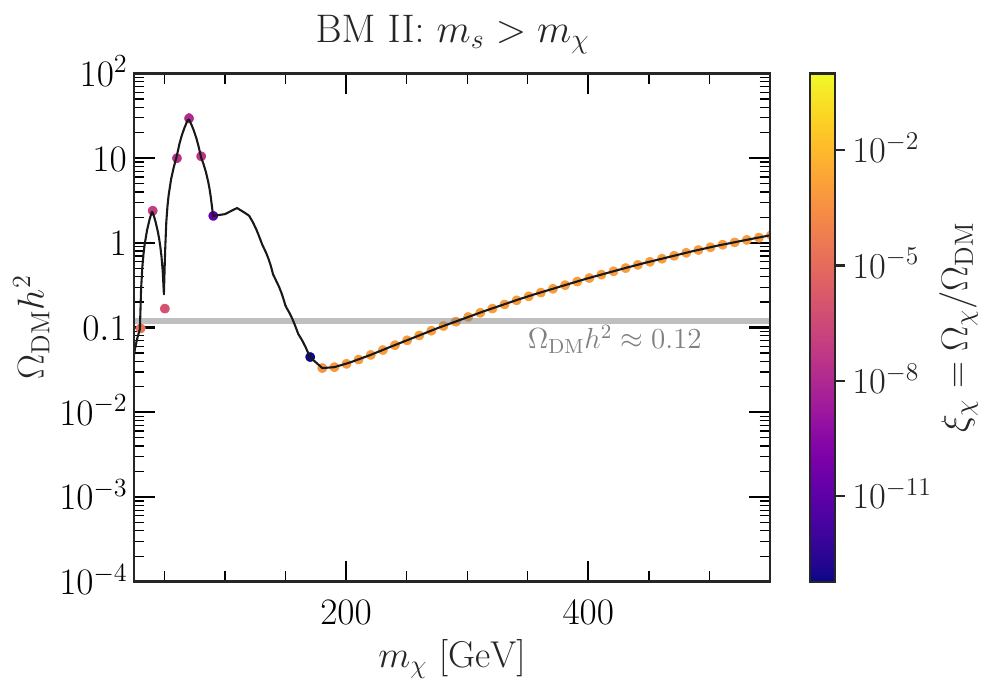}
    \caption{We plot $\Omega_{\rm DM}h^2 = (\Omega_\chi + \Omega_s)h^2$ as a function of the DM masses, $m_\chi$ or $m_s$. The horizontal gray band corresponds to the region compatible with PLANCK observations~\cite{Planck:2018vyg}. The colorbar indicates the fractional abundance of each component, $\xi_{\chi,s} = \Omega_{\chi,s}/\Omega_{\rm DM}$. The remaining parameters are fixed to the benchmark points defined in Tab.~\ref{tab:benchmarks}, obtained from the full parameter-space scan.}
    \label{fig:omegaplot}
\end{figure}

After solving this set of equations, the relic density is given by
\begin{align}\label{relic}
\Omega_i h^2 \simeq 2.83 \times 10^8
\, Y_{i,0} \, \frac{m_i}{\rm GeV} \; ,
\end{align}
where $Y_{i,0}$ is the yield of each DM component today, {\it i.e.}, after freeze-out, and $h$ is the reduced Hubble constant, $h\approx 0.67$. For illustrative purposes, in \autoref{fig:omegaplot}, we show the total DM abundance as a function of the DM masses, where the gray line corresponds to the PLANCK result, \autoref{eq:DMbudget}. We obtain the solutions by fixing the free parameters to the benchmark points I and II in \autoref{tab:benchmarks}. (See \autoref{sec:scan} for details on these benchmarks). We highlight the richness of the freeze-out mechanism due to the presence of two Higgs portals (in the resonance-induced valleys) and new final states in DM annihilation processes (at thresholds as the DM mass grows).

\section{Phenomenology}
\label{sec:pheno}

In this section, we discuss the relevant experimental constraints on the model and comprehensively scan its parameter space. After identifying points consistent with the empirical data, we scrutinize the model's viability in light of collider results.

\subsection{Experimental constraints}\label{sec:experiments}
\paragraph{Relic density}
We are interested in a two-component scenario. Thus, we require the sum of both DM abundances to satisfy the constraint reported by the PLANCK collaboration at $68\%$ C.L.~\cite{Planck:2018vyg}
\begin{equation}\label{eq:DMbudget}
0.1186 \leq (\Omega_s+\Omega_\chi) h^2 \leq  0.1236 \,. 
\end{equation}

\paragraph{Invisible Higgs decays}
If kinematically allowed, {\it i.e.}, $m_s < 125/2$ GeV, the SM-like Higgs boson $h$ or $H$ can decay into two scalar DM particles, contributing to the branching ratio of the Higgs into invisible particles. The corresponding decay width is
\begin{equation}
\Gamma_\text{inv}(h,H\to ss) = \frac{\lambda_{(h,H)s}\,v^2}{32\pi m_{h,H}} \, \sqrt{1-\frac{4m_s^2}{m_{h,H}^2}} \; ,
\end{equation}
in the heavy or light partner scenario, respectively.
To be consistent with the experimental data~\cite{ParticleDataGroup:2024cfk}, we require
\begin{equation}\label{eq:inv}
    \operatorname{Br}(h,H\to \text{inv})=\frac{\Gamma_\text{inv}}{\Gamma_\text{SM}+\Gamma_\text{inv}} < 0.107
\end{equation}
at $95\%$ C.L, where $\Gamma_\text{SM}=3.7^{+1.9}_{-1.4}$ MeV.

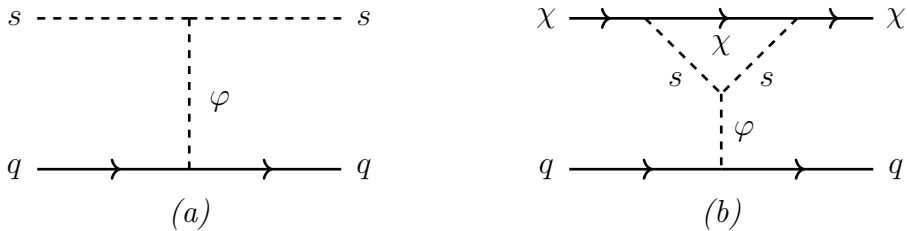
\begin{figure}[t]
\centering
\begin{tikzpicture}[line width=1.0 pt, scale=1.0]
\begin{scope}[shift={(0,0)}]
    % External scalars
    \draw[scalarnoarrow] (-2,1) -- (0,1);
    \draw[scalarnoarrow] (0,1) -- (2,1);
    % Fermions
    \draw[fermion] (-2,-1) -- (0,-1);
    \draw[fermion] (0,-1) -- (2,-1);  
    % Higgs propagator
    \draw[scalarnoarrow] (0,1) -- (0,-1);
    % Labels
    \node at (-2.3,1) {$s$};
    \node at (2.3,1) {$s$};
    \node at (-2.3,-1) {$q$};
    \node at (2.3,-1) {$q$};
    \node at (0.4,-0.1) {$\varphi$};
    \node at (0,-1.6) {$\textit{(a)}$};
\end{scope}
\begin{scope}[shift={(7,0)}]
    % Upper fermion line
    \draw[fermion] (-2,1) -- (-1,1);
    \draw[fermion] (-1,1) -- (1,1);
    \draw[fermion] (1,1) -- (2,1);
    % Loop
    \draw[scalarnoarrow] (-1,1) -- (0,0);
    \draw[scalarnoarrow] (1,1) -- (0,0);
    \draw[scalarnoarrow] (0,0) -- (0,-1);
    % Lower fermion line
    \draw[fermion] (-2,-1) -- (0,-1);
    \draw[fermion] (0,-1) -- (2,-1);
    % Labels
    \node at (-2.3,1) {$\chi$};
    \node at (2.3,1) {$\chi$};
    \node at (-2.3,-1) {$q$};
    \node at (2.3,-1) {$q$};
    \node at (0.3,-0.5) {$\varphi$};
    \node at (-0.6,0.2) {$s$};
    \node at (0.6,0.2) {$s$};
    \node at (0,0.65) {$\chi$};
    \node at (0,-1.6) {$\textit{(b)}$};
\end{scope}
\end{tikzpicture}
\caption{Feynman diagrams for DM--quark elastic scattering. (a) Tree-level process for scalar DM $s$. (b) One-loop process for fermionic DM $\chi$.}\label{dddiagrams}
\end{figure}

\paragraph{Direct detection}
Both DM candidates can scatter elastically off nuclei, providing a signature that is probed by direct detection experiments such as XENONnT~\cite{XENON:2023cxc}, LUX-ZEPPELIN (LZ)~\cite{LZ:2022lsv}, and the proposed DARWIN experiment~\cite{DARWIN:2016hyl}. In a multi-component DM scenario, the predicted cross-section must be rescaled by the fractional abundance of each component when comparing with experimental limits on the spin-independent cross-section:
\begin{equation}
\hat{\sigma}_\text{SI}^i=\frac{\Omega_i}{(\Omega_s+\Omega_\chi)}\sigma_\text{SI}^i, \quad i=s,\chi \; .
\end{equation}
The presence of the two Higgs fields, $h$ and $H$, and their couplings to $s$, allows elastic scattering between the scalar DM and nuclei. In diagram (a) of \autoref{dddiagrams}, we present the tree-level contributions to this process. On the other hand, elastic scattering of $\chi$ off nuclei is only possible at one-loop level, shown in diagram (b). The dominant contributions to the cross-sections for direct detection arise from diagrams in which the scalar mediator is the lighter CP-even state, $h$. If we neglect the contribution that come from the heavy mediator $H$, the relevant cross-sections are approximately
\begin{equation}
\sigma_\text{SI}^s\approx\frac{\lambda_{hs}^2 f_n}{4\pi} \frac{\mu_{sn}^2 m_n^2}{m_h^4 m_s^2} \; ,
\end{equation}
and
\begin{equation}
    \sigma_\text{SI}^\chi\approx \frac{1}{\pi} \frac{\mu^2_{\chi n} m_n^2 f_p^2}{m_h^4} \left[\lambda_{sh} \frac{|y_s|^2f(r_{s\chi})+|y_p|^2g(r_{s\chi})}{16\pi^2m_\chi}\right]^2 \; ,
     \label{eq:fermiondd}
\end{equation}
where $r_{s\chi}= m_s^2/m_\chi^2$ and
\begin{subequations}
\begin{eqnarray}
f(r_{s\chi}) &=& \frac{r_{s\chi}^2-5r_{s\chi}+4}{\sqrt{r_{s\chi}(4-r_{s\chi}})}\arctan\left(\frac{\sqrt{4-r_{s\chi}}}{\sqrt{r_{s\chi}}}\right) +\frac{1}{2}\left[ \,2-\left(r_{s\chi}-3\right)\ln(r_{s\chi}) \, \right] \; ,
\\
g(r_{s\chi}) &=& \frac{(r_{s\chi}-3)\sqrt{r_{s\chi}}}{\sqrt{4-r_{s\chi}}}\arctan\left(\frac{\sqrt{4-r_{s\chi}}}{\sqrt{r_{s\chi}}}\right)+\frac{1}{2}\left[\,2-\left(r_{s\chi}-1\right)\ln(r_{s\chi})\,\right] \; .
\end{eqnarray}
\end{subequations}

The crossed term $\sim y_s \, y_p$ in the 1-loop diagram is velocity-suppressed and, therefore, we ignore it for direct detection considerations~\cite{Yaguna:2021rds}.

\paragraph{Oblique corrections}
The states of the scalar sector of the 2HDM contribute to the self-energies of the SM gauge bosons at one-loop level, in contrast to the DM candidates $s$ and $\chi$, which are electroweak singlets. Peskin-Takeuchi oblique parameters $S$, $T$, and $U$ quantify contributions that are strongly constrained by experimental data. Concretely, we use
\begin{subequations}
\begin{eqnarray}
S|_{U=0} &=& 0.04\pm0.08 \; ,
\\
T|_{U=0} &=& 0.08\pm0.07 \; ,
\end{eqnarray}
\end{subequations}
with a correlation coefficient of 0.92~\cite{Haller:2018nnx}. Precision measurements of top quark and $W$ boson masses modify the determination of these values~\cite{deBlas:2022hdk, ParticleDataGroup:2024cfk}.
For explicit formulae regarding electroweak precision tests, see~\cite{Haber:2010bw,Funk:2011ad,Branco:2011iw}. For more general 2HDMs, see~\cite{Broggio:2014mna}.

\subsection{Numerical scan}\label{sec:scan}
In this section, we detail the numerical scan procedure that illustrates the main aspects of the model's phenomenology. We implement the model in \texttt{SARAH}~\cite{Staub:2009bi,Staub:2010jh,Staub:2012pb,Staub:2013tta,Porod:2014xia,Goodsell:2014bna,Goodsell:2015ira,Goodsell:2017pdq,Braathen:2017izn,Goodsell:2018tti}, and compute observables using  \texttt{SPheno}~\cite{Porod:2003um,Porod:2011nf}, \texttt{micrOmegas}~\cite{Belanger:2001fz,Belanger:2006is,Belanger:2010gh,Belanger:2013oya,Belanger:2014vza,Belyaev:2012qa,Alguero:2023zol}, and the analytic expressions shown on the previous section.
For our scan, we use log-uniform distributions to randomly generate values in the parameter space described in \autoref{eq:inputpar}. We generate mass parameters in the range $[1$ GeV, $1$ TeV$]$, imposing the constraint $m_s<2m_\chi$ to have a two-component DM scenario. Additionally, we generate quartic couplings such that their absolute values are in the range $[10^{-6},4\pi]$. This choice guarantees thermalization of DM candidates in the early Universe and imposes some perturbativity conditions. The angle  $\beta$ varies in the range $[10^{-3},\pi/2]$. After that we filter the sample to satisfy the remaining perturbativity conditions and the vacuum stability criteria. Finally, we numerically compute the cross-sections inferred from the diagrams in \autoref{diagrams} and we solve the Boltzmann equations~\autoref{boltzc} to obtain the relic density for both DM candidates, in order to keep only the points that fulfill the whole DM budget as indicated in \autoref{eq:DMbudget}. 

First, we analyze the behavior of the DM partition. As expected, the roles of semiannihilation and conversion are crucial. The endo- or exothermic nature of the processes described in \autoref{diagrams} depends on the mass hierarchy among the DM candidates and the portal states $h$ and $H$. Concretely, the semiannihilations (a) are open or closed depending on the sign of $2m_\chi-m_\chi-m_{h,H}$, while the semiannihilations (b) depend on the sign of $m_s-m_{h,H}$, as well as annihilations (f)-(h). The DM particle populated by conversion processes through contributions (c) and (d) depends on the sign of $m_s-m_\chi$. Finally, the possible final states of annihilations (e) depend on the mass hierarchy of the scalar particle with respect to all visible particles in the model.
To quantify the influence of these hierarchies on the freeze-out, in the following plots, we distinguish between the regimes of light and heavy Higgs partners, as well as those in which the scalar DM particle is lighter or heavier than the fermion DM. Also, we define the semiannihilation fractions to compare the impact of several processes~\cite{Yaguna:2021rds},
\begin{subequations}
\begin{eqnarray}
\zeta_s &\equiv& \frac{\frac{1}{2}\langle\sigma_{s\chi\chi\varphi}v\rangle}{\langle \sigma_{ssXX}v\rangle+\frac{1}{2}\langle \sigma_{s\chi\chi\varphi}v\rangle+\langle \sigma_{ss\chi\bar{\chi}}v\rangle} \; ,
\label{eq:zetas}    
\\
\zeta_\chi &\equiv& \frac{\frac{1}{2}\langle \sigma_{\chi\bar{\chi}s\varphi}v\rangle}{\frac{1}{2}\langle \sigma_{\chi\bar{\chi}s\varphi}v\rangle+\langle \sigma_{\chi\bar{\chi ss}} v\rangle} \; .
 \label{eq:zetachi}   
\end{eqnarray}
\end{subequations}

\begin{figure}[t]
    \centering
\includegraphics[width=0.49\linewidth]{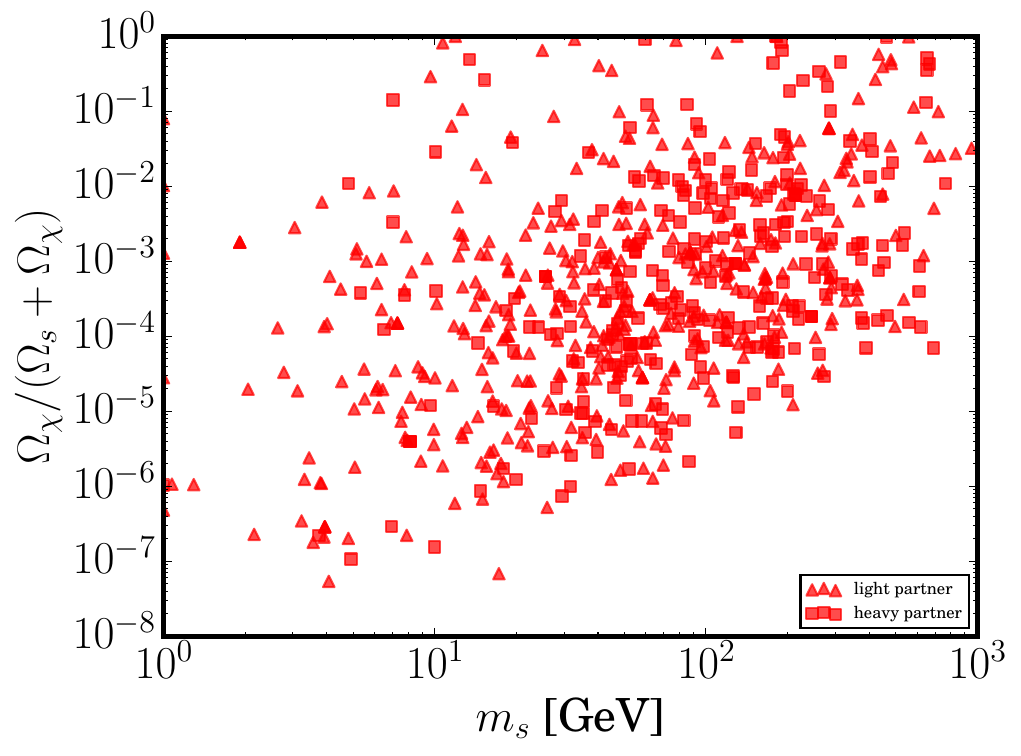}
\includegraphics[width=0.49\linewidth]{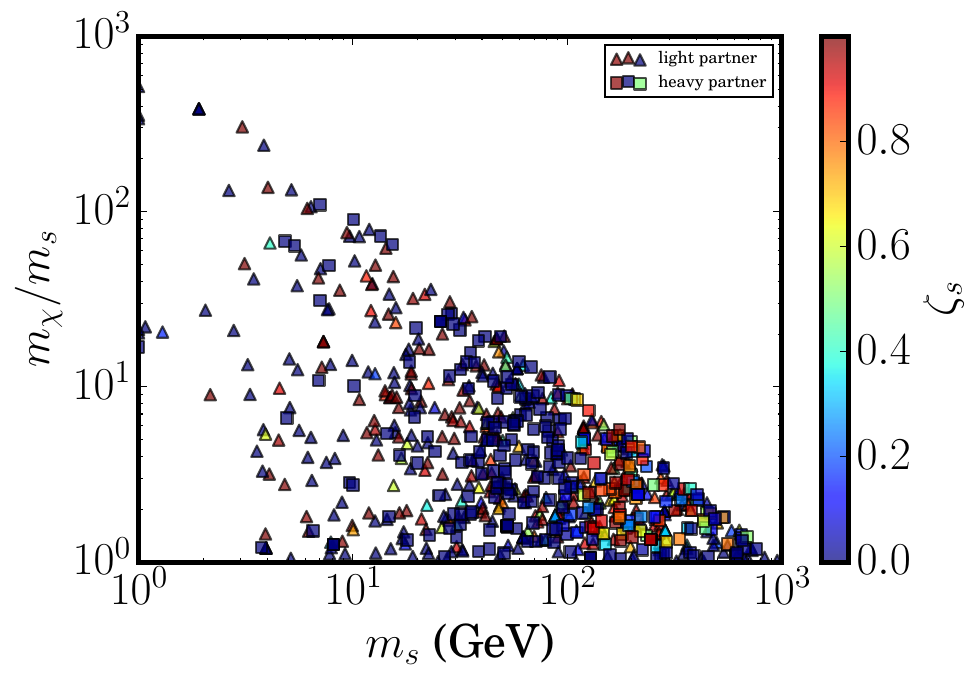}    
    \caption{Numerical scan satisfying $m_s < m_\chi$. Triangles denote points with $m_H = 125$ GeV (light partner regime), while squares denote points with $m_h = 125$ GeV (heavy partner regime). Left panel: projection onto the $m_s$ versus fermionic DM fraction plane. Right panel: projection onto the $m_s$ versus $m_\chi/m_s$ plane. The colorbar indicates the scalar semi-annihilation fraction $\zeta_s$, defined in \autoref{eq:zetas}.}
    \label{fig:lights}
\end{figure}

We project the numerical scan in \autoref{fig:lights} for the regime $m_s<m_\chi$. We emphasize that all points satisfy the PLANCK relic density constraint. The plot on the left shows the fraction of DM density for the fermion $\chi$. On the right plot, we show the ratio $m_\chi/m_s$ with colors representing the semi-annihilation fraction $\zeta_s$. We are interested in the DM relic abundance partitioning between scalar and fermion particles. From the left plot, we see that the fermion DM remains subdominant against the scalar DM, except for some particular points, independent of the regime of the Higgs partner, with the lowest fermionic fraction being of order $10^{-8}$ for $m_s \lesssim 100$ GeV. The larger the scalar mass, the lower the intensity of conversion from fermionic to scalar DM, so this minimum value gradually increases, and reaches $\sim 10^{-4}$ for $m_s\sim 1$ TeV.  
From the right plot, we see that the effect of semiannihilation is relatively weak compared with the effect of annihilation and conversion. We expect an increase in the scalar annihilation rate in a 2HDM compared to the SM extension in~\cite{Yaguna:2021rds} because a 2HDM model contains more final states in its scalar sector, in addition to the new Higgs portal. 

\begin{figure}[t]
    \centering
    \includegraphics[width=0.49\linewidth]{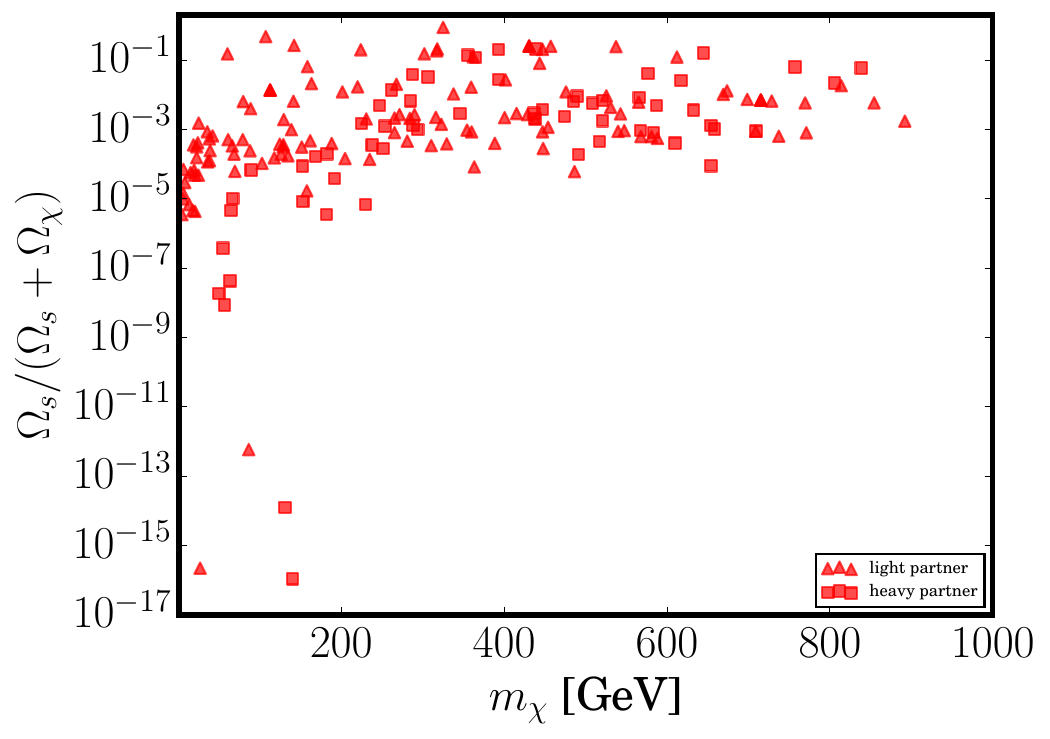}
    \includegraphics[width=0.49\linewidth]{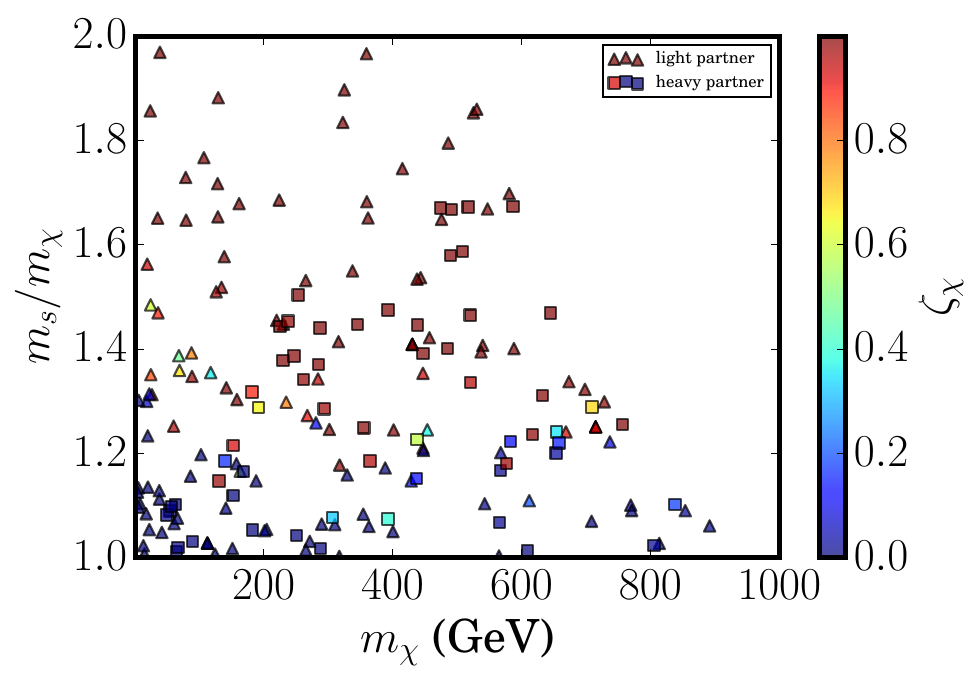} 
    \caption{Numerical scan satisfying $m_s > m_\chi$. Triangles denote points with $m_H = 125$ GeV (light partner regime), while squares denote points with $m_h = 125$ GeV (heavy partner regime). Left panel: projection onto the $m_\chi$ versus scalar DM fraction plane. Right panel: projection onto the $m_\chi$ versus $m_s/m_\chi$ plane. The colorbar indicates the fermionic semi-annihilation fraction $\zeta_\chi$, defined in \autoref{eq:zetachi}.}
    \label{fig:heavys}
\end{figure}
We project the numerical scan in \autoref{fig:heavys} for the regime $m_s>m_\chi$. The plot on the left shows the fraction of DM density that comes from the scalar $s$. Most points present scalar fractions of about $10^{-1}$ to $10^{-4}$ for $m_\chi$ up to 1 TeV. For $m_\chi \lesssim 200$ GeV, this fraction can decrease several orders of magnitude, with some points reaching $10^{-8}$.  Nevertheless, the scalar fraction can be higher at certain points, ranging from $10\%$ to $100\%$, specifically when a light Higgs portal is allowed.  
The presence of the two portals partially accounts for the overall rise of the scalar relic density in approximately one order of magnitude with respect to the SM extension studied in~\cite{Yaguna:2021rds}. In particular, those points with a light Higgs partner that account for more than $10\%$ of the scalar DM fraction undergo some degree of exponential increase of $Y_s$ before $s$ completely freezes out due to the semi-annihilation $\chi\bar{\chi} \to s \varphi$, a behavior known as \textit{bouncing DM}~\cite{Puetter:2022ucx, DiazSaez:2023wli}. This exponential growth, though typical of non-thermal freeze-in production, is specific to semi-annihilating multi-component DM models.
On the right plot, we show the ratio $m_s/m_\chi$ with colors indicating the semi-annihilation fraction $\zeta_\chi$. The mass of the scalar can be as heavy as twice the fermion mass, as long as $m_\chi \lesssim 700$ GeV. In that range, $\chi \bar{\chi}\to s \varphi$ is the dominant process in DM production, unless a degree of degeneracy with $m_s/m_\chi \lesssim 1.3$ is present, in which case annihilation and conversion processes are dominant. Also, notice that the light partner regime allows viable points in a region where the SM extension in~\cite{Yaguna:2021rds} excludes. 
In any case, we contrast these novelties with the strong constraints of direct detection experiments, which can exclude even points with a small scalar fraction, especially in the light Higgs partner regime.

\begin{figure}[t]
    \centering
    \includegraphics[width=0.75\linewidth]{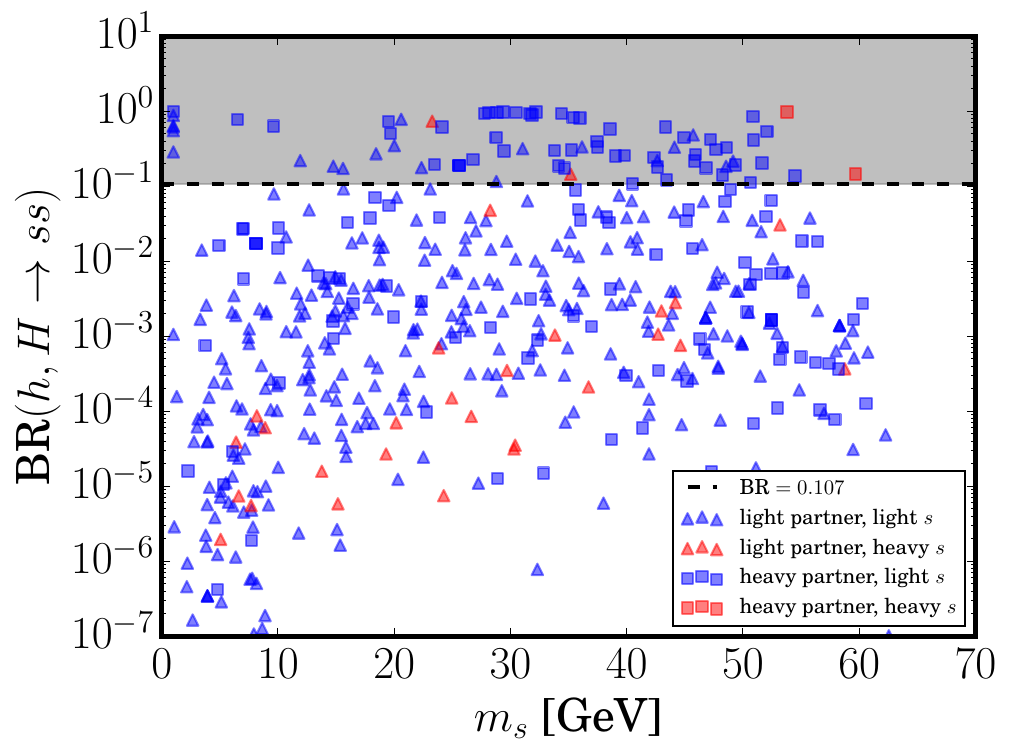}
    \caption{Numerical scan projected onto the invisible branching ratio of the SM-like Higgs boson as a function of the scalar mass for $m_s < 125/2$ GeV, computed using \autoref{eq:inv}. The horizontal line denotes the experimental upper bound.}
    \label{fig:invisible}
\end{figure}
In \autoref{fig:invisible}, we project the numerical scan onto invisible Higgs decays by evaluating \autoref{eq:inv}. This constraint excludes only points with $m_s<125/2$ GeV, so the most affected regime is $m_s<m_\chi$ (blue points). The relatively rare points with $125/2$ GeV $>m_s>m_\chi$ that can account for the correct DM relic density can easily bypass the upper bound for invisible Higgs decays. It is extremely difficult to satisfy the relic density with $125/2$ GeV $>m_s>m_\chi$ and a heavy Higgs partner, and those points seem to be completely excluded by constraints from invisible Higgs decays.

\begin{figure}[t]
    \centering
    \includegraphics[width=0.49\linewidth]{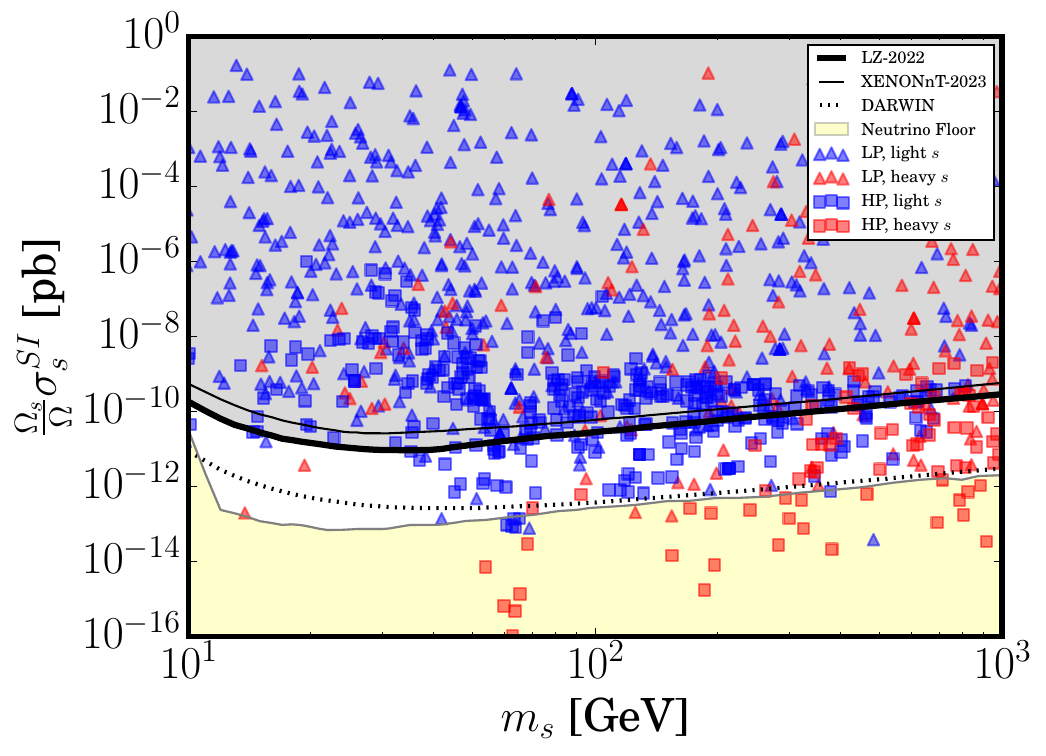}
    \includegraphics[width=0.49\linewidth]{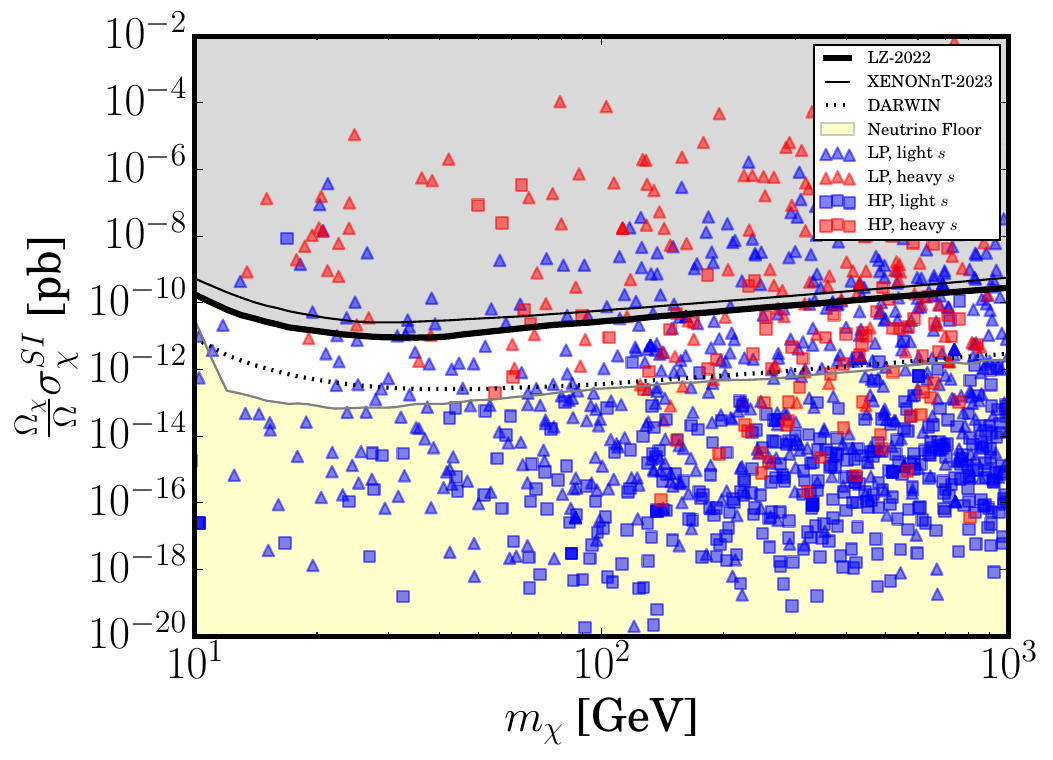}
    \caption{Numerical scan projected onto the spin-independent DM--proton cross-section as a function of the DM mass, rescaled by the fractional abundance of each component. We show current upper limits from XENONnT and LZ, as well as projected sensitivity from DARWIN. Left panel: tree-level scattering of scalar DM. Right panel: one-loop scattering of fermionic DM. }
    \label{fig:DD}
\end{figure}
In \autoref{fig:DD}, we plot the results for elastic scattering of DM with protons. In the left plot, we project the scalar DM signal, which we evaluate with~\texttt{micrOmegas}~\cite{Belanger:2001fz,Belanger:2006is,Belanger:2010gh,Belanger:2013oya,Belanger:2014vza,Belyaev:2012qa,Alguero:2023zol}, that automatically includes all diagrams (and their interference) in the calculation. In contrast, on the right plot, we use the analytic value given by \autoref{eq:fermiondd} for the fermion DM, which considers the lightest mediator $h$. 

From the plot on the left, we see that the direct-detection signal for the scalar DM candidate severely constrains this model. Although the four regimes can feature a sufficiently small signal, the regime with a light Higgs partner has obvious difficulties in evading the upper bounds due to the mediator mass. We also see that the regime with $m_s<m_\chi$, in which the relic density is typically dominated by $s$, is more constrained than the regime with heavy-fermion DM, since it lacks abundance suppression. At the end, the most suitable regime to bypass the bounds on these experiments is $m_s>m_\chi$ and a heavy Higgs partner.

From the right plot, we highlight that the direct detection signal of the fermion DM, even at one-loop level, has complementary constraining power, because the two regimes defined by the sign of $m_s-m_\chi$ present a relative dominance of one DM component over the other, so their direct detection signatures are suppressed in opposite ways. It is sensitive to a significant sample of the scanned points.  

\begin{figure}[t]
    \centering
    \includegraphics[width=0.75\linewidth]{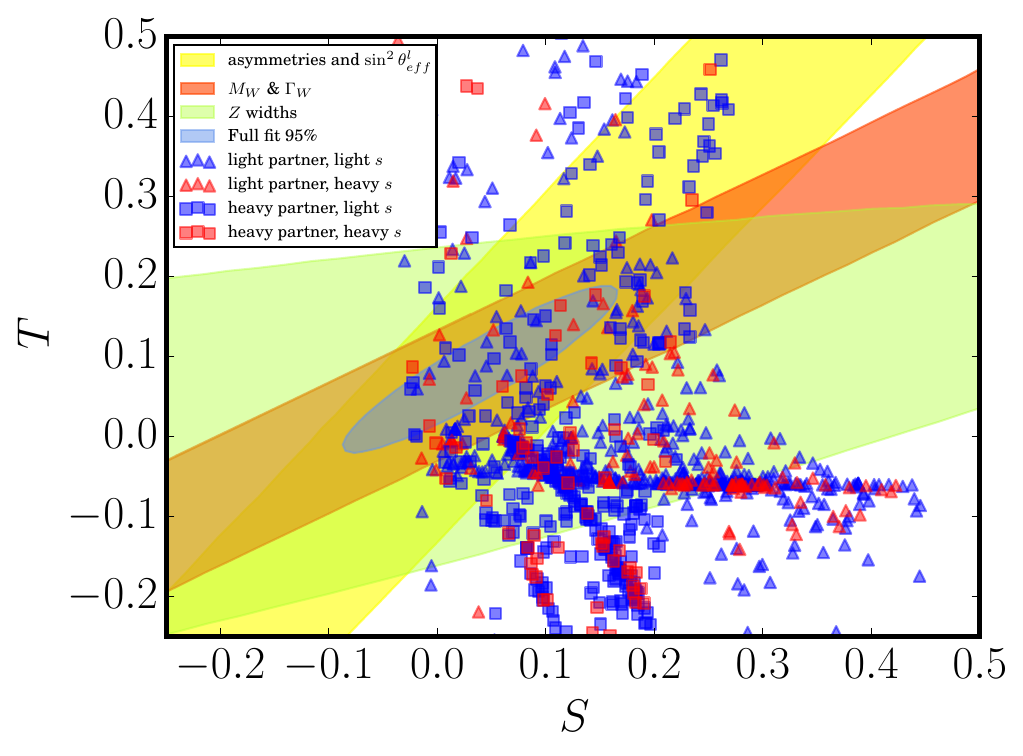}
    \caption{Numerical scan projected onto the $S$--$T$ plane. The blue and light-blue ellipses denote the $65\%$ and $95\%$ confidence level (C.L.) regions from a global fit. Constraints from asymmetry and direct $\sin^2\theta_{\text{eff}}^l$ measurements (yellow), the $Z$ partial and total widths (green), and the $W$ mass and width (red) are also shown.}
    \label{fig:STU}
\end{figure}

In \autoref{fig:STU}, we project the numerical scan onto the $S$ vs. $T$ plane. We calculate the numerical values of these quantities via \texttt{SPheno}~\cite{Porod:2003um,Porod:2011nf}. The plot clearly shows that electroweak precision tests impose severe constraints on the 2HDM. The high degree of degeneracy required in the visible scalar sector, for fixed mixing angles, is a serious flaw in the phenomenology of the 2HDM. It severely shrinks the set of experimentally consistent points generated by our scan procedure. As all scalar particles mediate one-loop self-energies for the gauge bosons, we find no clear correlation between the regimes defined in our scan and these observables. 

\begin{table}
\centering
\begin{tabular}{|c|c|c|c|c|}
\hline
  \textbf{Benchmark}   & \textbf{I}   & \textbf{II}  & \textbf{III}  & \textbf{IV}  \\ \hline \hline
$m_h$ & 125.10 & 125.10 & 72.69 & 122.53 \\
\hline
$m_H$ & 228.05 & 669.03 & 125.10 & 125.10 \\
\hline
$m_A$ & 168.36 & 1.09 & 251.33 & 70.26 \\
\hline
$m_{h^+}$ & 241.27 & 205.66 & 2.81 & 8.37 \\
\hline
$\beta$ & 1.21 & 1.02 & 0.83 & 0.14 \\
\hline
$\mu_{12}$ & 48.49 & 2.25 & 1.93 & 7.74 \\
\hline
$m_s$ & 538.74 & 378.03 & 650.19 & 270.92 \\
\hline
$m_\chi$ & 973.28 & 294.02 & 457.27 & 445.38 \\
\hline
$\lambda_{1s}$ & -0.167 & 1.115 & 0.258 & 0.0811 \\
\hline
$\lambda_{2s}$ & 0.0107 & -0.0118 & $-2.26\times10^{-5}$ & $-5.87\times10^{-4}$ \\
\hline
$g_L$ & -5.51 & 1.47 & -6.21 & 7.84 \\
\hline
$g_R$ & 8.10 & 1.83 & 6.06 & -1.35 \\
\hline
%$\lambda_s$ & 0.405 & 6.105 & $7.73\times10^{-5}$ & 0.176 \\
%\hline
\end{tabular}
\caption{Benchmark points satisfying perturbativity, vacuum stability, relic density, invisible Higgs decay, direct detection, and electroweak precision constraints. Masses are given in GeV.}
\label{tab:benchmarks}
\end{table}

We identify %one 
benchmark points for each regime (light or heavy Higgs partner, lighter or heavier scalar DM particle) by filtering our numerical scan against all experimental bounds discussed in \autoref{sec:experiments}, as shown in \autoref{tab:benchmarks}. In the following section, these benchmarks are contrasted with collider results for the scalar sector. It is worth noting that the intersection of exclusion regions from different experiments in the same parameter space generally reduces the confidence level of the intersection relative to that of individual experiments. A more correct approach consists of a combined likelihood analysis with $\chi^2$ functions summed in quadrature~\cite{AbdusSalam:2020rdj}.

\subsection{Collider signals}
We summarize here the constraints from collider searches on this extension of the type-I 2HDM. In principle, collider data can test this scenario through missing-energy signatures from DM production, such as \textit{mono-X} events, and through direct searches for the additional Higgs states in visible final states. In practice, for the benchmark points in \autoref{tab:benchmarks}, missing-energy searches have limited sensitivity in the DM-mass range of interest, so the dominant collider constraints arise from searches for the extended scalar sector.

\paragraph*{Flavor Physics}
Precision measurements of flavor-physics observables impose stringent constraints on the 2HDM scenarios, particularly in type-II and type-F models. In the type-I 2HDM, the limits are generally weaker, but charged Higgs bosons with masses up to about 900 GeV can be excluded for $\tan\beta \lesssim 1$~\cite{Li:2024kpd}, ruling out benchmark points III and IV.

\paragraph*{LEP searches}
Charged Higgs bosons were directly searched for at LEP via $e^{+} e^{-} \to h^{+}h^{-}$. Considering the decay $h^{+} \to \tau^{+}\nu_{\tau}$, the ALEPH collaboration excludes $m_{h^{\pm}} < 79.3$ GeV if the decay $h^{\pm} \to AW^{\pm}$ is kinematically forbidden, independently of BR$(h^{+} \to \tau^{+}\nu_{\tau})$~\cite{ALEPH:2002ftu}. LEP also constrains the sum of neutral scalar masses to $m_A + m_H \gtrsim 209$ GeV via $e^{+} e^{-} \to AH$, with $H \to b\bar{b}/\tau\tau$ and $A \to b\bar{b}/\tau\tau$~\cite{ALEPH:2006tnd, Kling:2016opi}.

\paragraph*{LHC searches}
Direct searches at the LHC now cover the full Run-2 dataset (typically $139$--$140~\mathrm{fb}^{-1}$ in ATLAS and $138~\mathrm{fb}^{-1}$ in CMS), and they constrain the scalar spectrum through complementary channels. For our purposes, it is useful to distinguish (i) fermionic final states, whose rates depend strongly on the Yukawa structure, and (ii) bosonic and cascade decays, whose sensitivity is often driven primarily by kinematics (mass splittings and open thresholds) and by the proximity to the alignment limit.

\emph{Neutral heavy Higgs bosons in fermionic channels}
Heavy neutral Higgs searches in $\tau^+\tau^-$ and $t\bar t$ provide robust constraints whenever the heavy states have appreciable couplings to SM fermions. ATLAS has performed a search for heavy neutral Higgs bosons decaying to $\tau^+\tau^-$ using the full Run-2 dataset~\cite{ATLAS:2020tautau}, while both ATLAS and CMS have dedicated searches for scalar/pseudoscalar resonances in $t\bar t$ final states~\cite{ATLAS:2024tt, CMS:2025tt}.

\emph{Charged Higgs and top interactions}
For $m_{h^\pm}<m_t$, top decays $t\to h^+ b$ yield strong bounds in several $h^\pm$ decay modes~\cite{ATLAS:2023Hcb}. For $m_{h^\pm}>m_t$, ATLAS has searched for $pp\to tbh^\pm$ with $h^\pm\to tb$ over $m_{h^\pm}\sim 200$ -- $2000$ GeV~\cite{ATLAS:2021Htb}. Moreover, if the spectrum allows for bosonic decays such as $h^\pm\to HW^\pm$, CMS sets direct limits in this channel using the full Run-2 dataset~\cite{CMS:2023HpmHW}.

\emph{Bosonic and cascade decays (Vh, VV, hh, ZH)}
Searches for resonances decaying to $Vh$ ($V=W,Z$) probe both spin-1 resonances and CP-odd scalars, setting limits on $\sigma\times\mathrm{BR}$ over a wide mass range that include the sub-TeV region of interest~\cite{ATLAS:2023Vh}. In addition, heavy Higgs searches in $ZZ$ and $WW$ final states constrain scalar resonances produced via gluon fusion or vector-boson fusion. Earlier Run-2 searches were reported by ATLAS and CMS in the $ZZ$ and $WW$ channels~\cite{ATLAS:2018sbw,ATLAS:2017tlw,CMS:2018amk,ATLAS:2017uhp,CMS:2019bnu}, and more recent analyses continue to extend the sensitivity in these modes~\cite{ATLAS:2021ZZ,CMS:2024ZZ4l,ATLAS:2022WWconf}. Whenever kinematically open, cascade decays such as $A\to ZH$ provide particularly sensitive probes of non-degenerate spectra. ATLAS and CMS have searched for $A\to Zh$ and related cascade channels in leptonic and semileptonic final states~\cite{ATLAS:2017xel,CMS:2019qcx,ATLAS:2020AZH,ATLAS:2024AZH}. Finally, heavy Higgs decays into pairs of SM-like Higgs bosons, $H\to hh$, are directly constrained by LHC searches for resonant Higgs-pair production. Early Run-2 results already placed limits on resonant $hh$ production~\cite{ATLAS:2019qdc}, while recent ATLAS combinations provide improved sensitivity with an explicit interpretation in the type-I 2HDM~\cite{ATLAS:2024resHH}.

Taken together, the breadth of Run-2 LHC searches makes it challenging to accommodate a fully light scalar spectrum with $m_A,m_H,m_{h^\pm}$ in the sub-TeV region unless the parameters simultaneously sit close to alignment and features (approximate) mass degeneracies that close the most sensitive cascade channels, and/or have Yukawa-suppressed production rates at the relevant masses.
\section{Discussion and conclusion}
\label{sec:conclusions}

In this work, we have investigated a two-component DM scenario within an extension of the type-I two-Higgs-doublet model, where the stability of the dark sector components is ensured by a $Z_4$ symmetry and kinematic requirements. The model features a scalar and a fermionic singlet that interact through Higgs portals and Yukawa couplings, leading to a rich thermal history involving annihilation, conversion, and semi-annihilation processes.

Our numerical analysis shows that viable regions of parameter space exist that simultaneously reproduce the observed relic abundance and satisfy bounds from invisible Higgs decays and direct detection. However, once electroweak precision observables and collider constraints are imposed, the allowed parameter space is drastically reduced. In particular, we find a clear tension between the regions favored by DM phenomenology and those compatible with collider data.
Our results indicate that the sub-TeV regime, which is typically preferred by WIMP-like DM scenarios, is strongly constrained. The electroweak precision tests impose a high degree of mass degeneracy among the scalar states, while direct searches at colliders impose lower bounds on charged and neutral Higgs masses. As a consequence, most viable points in this region require a delicate balance between these constraints.

One possible way to bypass collider constraints is simply to consider heavier scalar spectra. In this regime, collider constraints become weaker, and the relic density can be readily accommodated. Additionally, direct detection constraints are relaxed as the DM mass increases, and the heavier masses relatively suppress contributions to oblique parameters in the loop propagators. However, this comes at a high cost. Maintaining a SM-like Higgs boson at 125 GeV while pushing the rest of the scalar spectrum to higher scales requires increasingly large quartic couplings in the scalar potential. This quickly leads to a breakdown of perturbativity or, at best, to a highly fine-tuned parameter space. 
From this perspective, the sub-TeV region remains the most natural regime of the model. This reinforces the conclusion that the model is under significant pressure from current data.

At the model-building level, we now comment on two possible directions to relax these constraints. First, alternative Yukawa structures of the 2HDM (such as the type-II, X, or F) can modify the DM phenomenology by altering the couplings to fermionic final states, affecting both the relic density and direct detection predictions. Moreover, some collider bounds are model-dependent and may be weakened in different realizations of the Yukawa sector. However, important collider tests, electroweak precision constraints, and perturbativity conditions remain largely insensitive to the specific 2HDM type. A second possibility is to depart from the standard thermal history. Non-standard cosmological scenarios or alternative production mechanisms could modify the relic abundance calculation, potentially opening regions of parameter space that are otherwise excluded under the freeze-out assumption.

In summary, the interplay between collider constraints, electroweak precision observables, and DM requirements places this class of models under significant tension. While viable regions exist, they are either highly constrained or rely on some degree of fine-tuning or non-standard assumptions. Future improvements in direct detection sensitivity and collider searches for extended Higgs sectors will be crucial in further testing the viability of this framework.
\section*{Acknowledgments}
We thank Diego Cogollo, Carlos Yaguna, Farinaldo Queiroz, Josef Pradler, David Cabo-Almeida, and Sven Fabian for thoughtful discussions.
This work is supported by Simons Foundation (Award Number:1023171-RC), FAPESP Grant 2018/25225-9, 2021/01089-1, 2023/01197-4, ICTP-SAIFR FAPESP Grants 2021/14335-0, CNPQ Grants 403521/2024-6, 408295/2021-0, 403521/2024-6, 406919/2025-9, 351851/2025-9, ANID-Millennium Science Initiative Program ICN2019\_044, and IIF-FINEP grant 213/2024.
JPN thanks the Federal Ministry of Women, Science and Research (BMFWF) via the \textit{Ernst Mach Grant, weltweit}, reference number MPC-2025-01008. JPN also thanks the National Council for Scientific and Technological Development (CNPq) under the grant number 307130/2021-5. JPN thanks the Brazilian Federal Agency for Support and Evaluation of Graduate Education (CAPES) under the grant number 88887.712383/2022-00. JPN thanks the Marietta Blau Institute for Particle Physics (MBI/ÖAW) for the hospitality during the development of this project. 
P. E. is supported by CNPq grant No. 151612/2024-2. 
C.R. is supported by CAPES projects no. 88887.645500/2021-00
and 88887.935477/2024-00. 
\appendix
\section{Mass matrices, eigenstates and mixing angles}\label{App:matrices}
In this appendix, we present the diagonalization of the scalar spectrum in detail.

After applying the minimization conditions, the mass matrix of the CP-even states is 
\begin{eqnarray}
M_{h,H}^{2}=
\left(
\begin{array}{cc}
\lambda_1v_1^2+\mu_1^2\frac{v_2}{v_1} & -\mu_{12}^2+\lambda v_1v_2 \\
-\mu_{12}^2+\lambda v_1v_2 & \lambda_2v_2^2+\mu_{12}^{2}\frac{v_1}{v_2} \\
\end{array}
\right),
\end{eqnarray}
and it is diagonalized by the orthogonal transformation
\begin{eqnarray}\label{transfh}
\left(
\begin{array}{c}
h \\
H \\
\end{array}
\right)
=\left(
\begin{array}{cc}
\cos\alpha & \sin\alpha \\
-\sin\alpha & \cos\alpha \\
\end{array}
\right)
\left(
\begin{array}{c}
\rho_1 \\
\rho_2 \\
\end{array}
\right).
\end{eqnarray}
where the angle $\alpha$ is given in \autoref{eq:alphadef}. The physical eigenstates $h$ and $H$ have masses given by \autoref{h_and_H}.

On the other hand, the mass matrices of the CP-odd and the charged states are given by
\begin{eqnarray}
M_{A}^{2}=
\left(\mu_{12}^2-\lambda_5v_1v_2\right)\left(
\begin{array}{cc}
\frac{v_2}{v_1} & -1 \\
-1 & \frac{v_1}{v_2} \\
\end{array}
\right),
\end{eqnarray}
and
\begin{eqnarray}
M_{h^{\pm}}^{2}=
\left(\mu_{12}^2-\frac{1}{2}(\lambda_4+\lambda_5)v_1v_2\right)\left(
\begin{array}{cc}
\frac{v_2}{v_1} & -1 \\
-1 & \frac{v_1}{v_2} \\
\end{array}
\right),
\end{eqnarray}
and these two matrices are diagonalized by the same orthogonal transformation:
\begin{eqnarray}
\left(
\begin{array}{c}
G \\
A \\
\end{array}
\right)
=\left(
\begin{array}{cc}
\cos\beta & \sin\beta \\
-\sin\beta & \cos\beta \\
\end{array}
\right)
\left(
\begin{array}{c}
\eta_1 \\
\eta_2 \\
\end{array}
\right),
\end{eqnarray}
and
\begin{eqnarray}
\left(
\begin{array}{c}
G^+ \\
h^+ \\
\end{array}
\right)
=\left(
\begin{array}{cc}
\cos\beta & \sin\beta \\
-\sin\beta & \cos\beta \\
\end{array}
\right)
\left(
\begin{array}{c}
\phi_1^+ \\
\phi_2^+ \\
\end{array}
\right).
\end{eqnarray}
where $\beta$ is given in \autoref{eq:betadef}, $G$ and $G^+$ are the Goldstone bosons absorbed by the $Z$ and $W^+$ bosons, and the masses of the physical $A$ and $h^+$ are given by \autoref{eq:Amass} and \autoref{eq:hpmass}.

\bibliographystyle{JHEPfixed.bst}
\bibliography{references}

\end{document}